\documentclass[aps,floats,superscriptaddress]{revtex4}
\usepackage{amsmath}
\usepackage[a4paper,top=3cm,bottom=2cm,left=1.6cm,right=2cm]{geometry}
\newcommand{\beq}{\begin{equation}}
\newcommand{\eeq}{\end{equation}}
\newcommand{\beqa}{\begin{eqnarray}}
\newcommand{\eeqa}{\end{eqnarray}}\usepackage{epsfig}
\usepackage{graphicx}
\usepackage{dcolumn}
\usepackage{bm}
\usepackage{slashed}
\begin{document}

\bibliographystyle{apsrev}
%
\author{Alberto Cortijo}
\affiliation{Department of Physics, Lancaster University, \\ Lancaster, LA1 4YB, United Kingdom.}
\title{Topological insulating phases in mono and bilayer graphene: an effective action approach}
\author{Adolfo G. Grushin}
\affiliation{Instituto de Ciencia de Materiales de Madrid,\\
CSIC, Cantoblanco, E-28049 Madrid, Spain.}
\author{Mar\'ia A. H. Vozmediano}
\affiliation{Instituto de Ciencia de Materiales de Madrid,\\
CSIC, Cantoblanco, E-28049 Madrid, Spain.}
\date{\today}

\begin{abstract}
We analyze the influence of different bilinear terms giving rise to time reversal invariant
topological insulating phases in mono and bilayer graphene. We make use of the effective action formalism to determine the dependence of the Chern Simons coefficient on the different interactions and generalize the formalism to include various degrees of freedom arising in bilayer graphene which give rise to different topological phases.

\end{abstract}

\maketitle

\section{Introduction}
Since its recent synthesis, graphene {\cite{Netal05,ZTSK05}
has become an ideal playground to establish fundamental concepts in condensed matter physics \cite{G09}. It is by now well known that the low energy excitations
of the neutral system can be described by a continuum model that includes
two flavors of massless Dirac particles in two spatial
dimensions \cite{RMP08}. The two flavors are associated to the two Fermi points
that constitute the Fermi surface of the neutral material and,
in the standard description, the electron spin is an extra degree of freedom
usually disregarded as long as the physics in the system
does not explicitly depend on it.
In the last years it has been realized that when
the physics involves the spin degree of freedom, interesting
things might happen and new states of matter can emerge  like the
topological insulating state \cite{KM05,KM05b,BHZ06,HK10}. This state relies on the existence
of a nonzero spin orbit (SO) coupling which is
believed to be very small in graphene \cite{KM05,MHetal06} but which recent
proposals show that it can be enhanced by various factors such as curvature \cite{CLM09,HGB06}, impurities \cite{CG09} or Coulomb interactions \cite{GGB10}. Also the application of a perpendicular
electric field in graphene can induce an extrinsic
Rashba SO coupling whose value can  also be enhanced by interactions
with the substrate \cite{VSetal08}. Apart from the SO processes it is
known that a staggered potential can open a gap in the
spectrum, existing some experimental evidence about the
gap opening process when graphene is deposited on top
of SiC \cite{Zetal07}. The way in which those terms compete among each other
can lead to a rich phase diagram of insulating and gapless
phases as discussed in \cite{Retal09} for monolayer graphene.

There are other ways to get a nontrivial insulating
state in graphene without invoking the presence of a
SO coupling. Recently it has been shown
theoretically that a proper choice of strain pattern can
lead to the appearance of a gapped phase,
with an energy spectrum similar to the system of Landau
levels and a pair of counterpropagating edge states,
carrying different value of the valley index \cite{GKG10}. Contrary to
what happens when the SO coupling is considered, here
we cannot establish that these counter-propagating states
are protected by some symmetry (when the SO coupling
is considered the edge states are protected against disorder by
time reversal symmetry). Nevertheless,
the topological character of the counter--propagating
states is maintained for distances smaller that the mean
free path in the system. Other situation where  a topological insulating phase can appear without a SO interaction is by deposition of  ad--atoms on the surface of graphene. When these ad--atoms are placed only in one of the two sublattices, a gap is opened and again, a pair of counter-propagating states appear at the boundaries carrying opposite valley number \cite{CSetal10}.

The goal of this work is to describe how time reversal invariant topological phases can be identified and analyzed in a unifying way using the effective action formalism, a technique which has been widely used in condensed matter systems \cite{F91}. We derive a way to compute the Chern-Simons coefficient associated to the different situations allowed by the bilinear terms in the hamiltonian. We will also see how terms different from the usual spin orbit coupling  can
lead to a Chern-Simons term even in the absence of an external electromagnetic field.
It is well known that  in the graphene system a finite density of carriers can be induced by applying an external voltage.   We will pay special attention to  the situation in which there is a  finite chemical potential in the samples. As we will see this changes some physical situations and introduce phases in which the system is metallic with a finite value of the Chern-Simons coefficient.

Bilayer graphene \cite{MF06} is a more promising material than the monolayer for potential applications due to the  possibility of opening a gap by applying an external  voltage \cite{Cetal07}.  Its possible insulating phases are also richer and are being subject of great attention recently  \cite{MBM08,GM10,G10,PSetal10}. We extend the formalism to the bilayer case and provide a table of the insulating phases arising from the competition of the various spin-orbit couplings discussed in \cite{G10}.

The article is structured as follows: In section \ref{sec_SCS} we  describe the model for the monolayer graphene used in this work and present the formalism of the effective action to derive an expression to compute the Chern Simons coefficient. In section \ref{sec_monolayer}  the formalism is  applied to study various insulating phases in the monolayer case with a finite chemical potential. Section \ref{sec_bil} is devoted  to analyze various non-trivial topologically insulating phases in bilayer graphene. We  conclude with a summary and discussion of the results in section \ref{sec_final}.

\section{ The spin Chern Simons coefficient from the effective action}
\label{sec_SCS}

In this section we review some of the known low energy aspects of graphene in the context of the effective action formalism for completeness. For details, we refer the reader to reference \cite{Retal09} and to the references mentioned below.

\subsection{Monolayer graphene: The model}
\label{sec_mono}
The low energy elementary excitations of neutral graphene around its Fermi points can be described by the following massless Dirac hamiltonian:
\begin{equation}
{\cal H}_{0}(\textbf{k})=i[\sigma_{y}\tau_{z}k_{x}-\sigma_{x}1_{\tau}k_{y}]1_{s}\equiv \gamma^{1}k_{x}+\gamma^{2}k_{y}.\label{Freeham}
\end{equation}
The Pauli matrices labeled $\sigma$, $\tau$ and $s$ represent pseudospin, valley and spin degrees of freedom respectively. The $\gamma$ matrices are  8x8 matrices  chosen  to fulfill the condition $\{\gamma^{\mu},\gamma^{\nu}\}=4\times2g^{\mu\nu}$, with $g^{\mu\nu}=\mathrm{diag}(1,-1,-1)$. With these conventions, the zeroth gamma matrix is defined as $\gamma^{0}=1_{s}1_{v}\sigma_{z}$. Since we will not consider many body effects  the Fermi velocity is set to one by a proper scaling of the fields and the parameters in the hamiltonian. We have also put $\hbar=1$. Note that the matrices chosen differ from the convention used in \cite{KM05} ($\gamma_{KM}$).  The two are related by $\gamma^\mu=\sigma_z\gamma^\mu_{KM}$.
 In what follows we will group all possible bilinear couplings with an arbitrary matrix $\Lambda$ and write the corresponding Lagrangean as
\begin{equation}\label{barelag}
\mathcal{L}_{0} = \bar{\psi}(\gamma^{\mu}k_{\mu}-\sum_{i}\lambda_{i}\Lambda_{i})\psi ,
\end{equation}
where $\bar{\psi}=\psi^+\gamma_0$. While each 2x2 Hamiltonian around a given Fermi point is not time reversal invariant ( $\cal{T}$: $H(K_1)=H^*(K_2)$ ) grouping the two valleys in a 4x4 notation makes the full Hamiltonian $\mathcal{H}_{0}$ invariant under both inversion symmetry (realized as an interchange of sub-lattices $A$ and $B$ often called parity $\mathcal{P}$ in the graphene context ) and time reversal $\mathcal{T}$(interchange of $K_1$ and $K_2$ ). These discrete symmetries --together with translation invariance--  protect the Fermi points and prevent the opening of a gap in the spectrum when the intrinsic spin of the electron is not taken into account \cite{MGV07}.

A staggered potential is a  time reversal invariant contribution (although it is a Dirac mass term that breaks parity) that opens a gap in the spectrum. In our notation it will  will be represented by a coupling constant $m$ followed by the unit matrix
\beq
\Lambda_{m}=1_{s}1_{\tau}1_{\sigma}.
\label{M}
\eeq

The main contribution of \cite{KM05} was to realize that when the spin degree of freedom is taken into account the intrinsic SO term acts like two copies of the Haldane mass \cite{H88}, with opposite sign for each spin projection. The Haldane mass was one of the first attempts to get Landau levels with net zero magnetic field \cite{S84}. With our convention it is defined as $\Lambda_{H}=1_{s}\tau_{z}1_{\sigma}$ and breaks $\cal{T}$.  The symmetry  is restored by the inclusion of the two opposite spin contributions.
In our convention the intrinsic SO coupling  has the form:
\begin{equation}\label{SOham}
H_{so}=\Delta_{so}\Lambda_{so}\equiv \Delta_{so}s_{z}\tau_{z}1_{\sigma}.
\end{equation}
We will also consider in our work the Rashba term which is written as
\begin{equation}\label{Rham}
H_{R}=\lambda_{R}\Lambda_{R}\equiv \lambda_{R}i(s_{y}\tau_{z}\sigma_{y}+s_{x}1_\tau\sigma_{x}).
\end{equation}
As mentioned before, the SO term  opens a gap in the system and respects both time reversal and parity symmetry. The other time reversal invariant term that opens a gap is a spatially constant Kekul\'{e} distortion  which  is off-diagonal in the valley and lattice indexes. Its contribution to the Chern-Simons term discussed in this work is the same as that of the staggered potential and we will not consider it here.

\subsection{The spin Chern-Simons term}
It is well known that in $(2+1)$ space time dimensions, a single  massive Dirac fermion coupled to a background U(1) electromagnetic (EM) field  induces a topological Chern-Simons term in the effective action of the gauge field \cite{DJT82,R84}. In condensed matter language and in connection with the physics of the quantum Hall effect \cite{J84,I84,S84}) this can be seen as follows:
The action for the minimal model of a Dirac fermion $\psi$ coupled to an EM field described by the gauge potential $A_\mu$ is given by
\beq
S=\int d^3 x \left[\bar{\psi} \gamma^\mu(\partial_\mu +ieA_\mu)\psi +m\bar{\psi}\psi
-\frac{1}{4}F^{\mu\nu}F_{\mu\nu}\right].
\eeq
The first radiative correction to the vacuum polarization (related by the Kubo formula to the conductivity tensor) is given by
\beq
\Pi_{\mu\nu}(q)=ie^2\int \frac{d^3 k}{(2\pi)^3}\mathrm{Tr}\left[\gamma^\mu
\frac{1}{\gamma^\rho k_\rho -m}\gamma^\nu
\frac{1}{\gamma^\rho (k+q)_\rho -m}\right].
\eeq
From the properties of the Dirac matrices it is known that the vacuum polarization tensor can be decomposed as
\beq
\Pi_{\mu\nu}(q)=\left(g_{\mu\nu}-\frac{q_\mu q_\nu}{q^2}\right)\Pi^e(q^2)+
i m \epsilon_{\mu\nu\rho}q^\rho\Pi^o(q^2),
\label{poltensor}
\eeq
where the symmetric (antisymmetric) part  is even (odd) under inversion and time reversal symmetries.   The tensor structure of the odd term, which emerges from the
trace of three $\gamma$ matrices, induces a Chern-Simons interaction in the effective action of the  electromagnetic field of the form
\begin{equation}
S_{cs}=C\int d^{3}x \epsilon^{\mu\nu\rho}A_{\mu}\partial_{\nu}A_{\rho},\label{CSaction}
\end{equation}
and  a transverse fermionic current:
\begin{equation}
<J^{\mu}> = C\epsilon^{\mu\nu\rho}\partial_{\nu}A_{\rho},\label{transcurrent}
\end{equation}
where $C$ is the Chern-Simons coefficient which is proportional to the sign of the fermionic mass.

When two Dirac fermions related by time reversal symmetry are considered, each of them will contribute to  C with an opposite value giving a null transverse current (\ref{transcurrent}). Time reversal symmetry dictates that $C_{\uparrow}=-C_{\downarrow}$ and hence the total Chern number is $C=C_{\uparrow}+C_{\downarrow}=0$. 

So in order to generate a Chern-Simons-like term in a time reversal invariant fashion we need  to generate different masses (open a gap in the spectrum)  for different members of the same pair of flavors.

In graphene there are four flavors of massless Dirac fermions related by pairs (valley number) by time reversal symmetry. To describe an insulating state originated by some mass like term, we need to couple the Dirac fermions to a proper background field which can describe the difference between a standard insulating phase and a topologically insulating phase. Given a flavor, say spin, we must couple the fermions to an external background $V_{\mu}$ field that sees the spin:
\begin{equation}
S=\int \frac{d^{3}k}{(2\pi)^3}[\bar{\psi}(\gamma^{\mu}k_{\mu}-\Lambda)\psi-\bar{\psi}\gamma^{\mu}\psi A_{\mu}-\bar{\psi}\gamma^{\nu}\gamma_{5}\psi V_{\nu}].\label{SCSfermionaction}
\end{equation}
The matrix $\gamma_5$ has to be chosen appropriately to  do the job and it should not be identified with the usual product of all the other matrices. Its role is to ensure that the traces in the calculation of the gauge effective action are nonzero (the details are explained in the next section).
For instance, in the case of the spin current discussed in \cite{KM05} the matrix $\gamma_{5}$  is defined as $\gamma_{5}=s_{z}1_{v}1_{\sigma}$. In this way after integrating out the fermions and calculating the lowest order terms in the effective action (see next section) , eq.  (\ref{transcurrent}) will be properly modified to give
\begin{equation}
<J^{\mu}_{\uparrow}-J^{\mu}_{\downarrow}>\equiv \frac{\delta S_{cs}}{\delta V_{\mu}}=C_{s}\epsilon^{\mu\nu\rho}\partial_{\nu}A_{\rho}\label{spinresponse}.
\end{equation}
The spin Chern-Simons coefficient $C_{s}$ is  defined as $C_{s}=C_{\uparrow}-C_{\downarrow}$ \cite{KM05} and it will be related to the electron's Green's function in the next section. As mentioned before, time reversal symmetry dictates that the total Chern number is zero but we can still find a transverse current being topological in origin given by $C_{s}$. This number classifies the quantum spin hall state. \\
This approach can be generalized to other degrees of freedom such as the valley degree of freedom by properly modifying $\gamma_{5}$. We will discuss this issue in the section devoted to the bilayer case. \\

\subsection{The effective action}
\label{sec_eff}

We will  describe in this section how to construct the spin Chern-Simons coefficient (or any other topological charge) on practical grounds when a generic mass like term $\Lambda$ is considered in the Lagrangean (\ref{barelag}), leaving the particular cases for the following sections. A similar construction was done in \cite{Setal08} in order to classify the topological insulators in three spatial dimensions and in \cite{Retal09} for monolayer graphene.
Consider the action defined in (\ref{SCSfermionaction}). The one-loop effective action for the gauge field is obtained by performing the functional gaussian integration of the fermionic variables. Since  (\ref{SCSfermionaction}) is quadratic in the fermions it is straightforward to get:
\begin{widetext}
\begin{equation}
\Gamma_{1}[A,V]=\int \frac{d^{3}k}{(2\pi)^{3}}\mathrm{Tr}\ln\left(\gamma^{\mu}k_{\mu}-\Lambda - e\gamma^{\mu}A_{\mu}-\gamma^{\nu}\gamma_{5}V_{\nu}\right).\label{effaction}
\end{equation}
\end{widetext}
It is important to note that in equation (\ref{effaction}) the momentum variable $k_{\mu}$ actually represents the derivative operator, so in order to deal with the logarithm in (\ref{effaction}) we need to expand it in Taylor series of the potentials $A_{\mu}$ and $V_{\nu}$. For that, we will formally define the fermionic Green's function as
\begin{equation}
G(k)=\frac{i}{\gamma^{\mu}k_{\mu}-\Lambda},\label{Green}
\end{equation}
and we will write (\ref{effaction}) 
as:

\begin{equation}\label{effaction2}
\Gamma_{1}[A,V]=\int\dfrac{dk^3}{(2\pi)^3}\sum_{n=1}^{\infty}-\dfrac{1}{n}\mathrm{Tr}\left[(G_{0}(k,\Lambda)(-ie\gamma^{\mu}A_{\mu} -ie\gamma_{5}\gamma^{\nu}\gamma_{5}V_{\nu}))^n\right],
\end{equation}
where we have dropped out  an irrelevant (infinite) constant that does not depend on the fields $A$ or $V$. In order to extract the spin Chern-Simons term
we can concentrate in the $l=2$ term in this Taylor expansion. Moreover since we will consider $k$ in (\ref{effaction2}) as the derivative operator, we will seek for the terms in the expansion that will contain derivatives of the fields $A$ and $V$, particularly  the term $V\partial A$, as it can be seen from (\ref{spinresponse}). We can perform the calculation in momentum space where the $l=2$ term reads:
\begin{equation}
\Gamma^{l=2}_{1}= \dfrac{e^2}{2}\int \frac{d^{3}q}{(2\pi)^{3}}\frac{d^{3}k}{(2\pi)^{3}}\mathrm{Tr}\left[G(k)\gamma^{\mu}G(k+q)\gamma^{\nu}\gamma_{5}\right]V_{\nu}A_{\mu}.\label{Feynmandiagram}
\end{equation}

Expanding $G(k+q)$ in series of $q$ up to first order we can identify the corresponding Chern Simons term which in real space reads:
%
%
\begin{equation}
S_{scs}=i\int d^{3}xC^{\mu\nu\rho}_{s}V_{\mu}\partial_{\nu}A_{\rho},
\end{equation}
where $C^{\mu\nu\rho}_{s}$ is defined as

%
%
\begin{eqnarray}\label{C5}
C_{s}^{\mu\nu\rho} = -i\dfrac{e^2}{2}\int \dfrac{dk^3}{(2\pi)^3} \mathrm{Tr}\left[G(k,\Lambda)\gamma^{\mu}G(k,\Lambda)\gamma^{\nu}G(k,\Lambda)\gamma^{\rho}\gamma_{5}\right].
\end{eqnarray}

It is interesting to note that the the last expression can be written in the form of a Pontryagin index \cite{V03,Y89}:
\begin{equation}
C^{\mu\nu\rho}_{s}=-i\dfrac{e^2}{2}\int \frac{d^{3}k}{(2\pi)^3}Tr [G\partial_{k^{\mu}}G^{-1}G\partial_{k^\nu}G^{-1}G\partial_{k^\rho}G^{-1}\gamma_{5}].\label{Pontryagin}
\end{equation}
In what follows  we will use these definitions to calculate some particular cases including the effect of the chemical potential.

\section{Topological insulating phases in monolayer graphene}
\label{sec_monolayer}

\subsection{An illustrative case: Intrinsic spin-orbit coupling}
\label{sec_SO}

Although known in the literature for quite some time \cite{KM05} it is illustrative to consider the case with only intrinsic spin orbit coupling as a starting point  to construct  more elaborate situations. In the simple case of the SO coupling with the chemical potential ($\mu$) set to zero, it is trivial to check that $\Lambda_{so}$ commutes with all the $\gamma$ matrices and thus we can write $G(k)$ as

\begin{equation}\label{easyGreen}
G(k)=i\frac{\gamma^{\rho}k_{\rho}+\Delta_{so}\Lambda_{so}}{k^{2}-\Delta^{2}_{so}}.
\end{equation}
Written $G(k)$ in this way, by the trace of the $\gamma$ matrices it is apparent that only the term proportional to $\Delta^{3}_{so}$ will survive, being proportional to $8i\epsilon^{\mu\rho\nu}$ as well. The integral in momentum can be evaluated easily, as long as we note that the integral is insensitive to the sign of $\Delta_{so}$. The result is \cite{KM05}:

\begin{eqnarray}
C_{s}^{\mu\nu\rho} = \epsilon^{\mu\nu\rho} \dfrac{e^2}{2}\int^{\infty}_{0}\dfrac{dk}{(2\pi)} \dfrac{2k\Delta_{so}}{(\mathbf{k}^{2}+\Delta_{so}^ {2})^{3/2}} = e^2\epsilon^{\mu\nu\rho} \frac{\Delta_{so}}{2\pi|\Delta_{so}|}= e^2\dfrac{\epsilon^{\mu\nu\rho}}{2\pi}\mathrm{sign}(\Delta_{so}).
\label{scscoeff2}
\end{eqnarray}
This is a widely known result. It can be proven that this number, $\mathrm{sign}(\Delta_{so})$ is the same $Z_{2}$ topological invariant defined by Kane and Mele in the case of two dimensional topological insulators \cite{KM05b}.\\
This result is modified by the presence of a finite chemical potential. When $|\mu|>\Delta_{so}$, the chemical potential crosses one of the bands changing the position of the poles in (\ref{C5}). Thus, we obtain:
\begin{eqnarray}\label{so+mu}
C_{s}^{\mu\nu\rho}=\epsilon^{\mu\nu\rho}\frac{\Delta_{so}e^2}{2\pi|\mu|},
\end{eqnarray}
a result first obtained in \cite{DDB09}. When $|\mu|<\Delta_{so}$ the two poles in (\ref{C5}) are located always on different semi-planes (upper and lower) and so the integral gives the same result as in the case of zero chemical potential. With this result we learn how to proceed with more general cases and the effect that we should expect with the introduction of a chemical potential. When the chemical potential crosses the bands, the topological nature of the effective action is not destroyed, but the result is no longer quantized and depends explicitly on the position of the chemical potential. When the chemical potential lies inside the gap, the system is a topological insulator with a quantized spin Chern-Simons response.

\subsection{The competition between the intrinsic spin-orbit term and the staggered potential}
\label{sec_SO-sta}
We will follow the method sketched in the previous section to study analytically the competition of the SO term with the staggered potential in the presence of a finite (and zero) chemical potential. The interest of this case resides on the fact that while both terms open a gap in the system, the nature of the insulating phase is different. It also introduces some technical complications whose resolution will be useful for the analysis of more complicated situations. The competition of  the intrinsic spin-orbit term and the staggered potential was studied numerically in \cite{KM05} for zero chemical potential. In what follows we will complete the discussion and give an analytical result valid for finite $\mu$. A similar analysis was also  done in \cite{Retal09} where the interplay of a Haldane mass and a staggered potential for zero chemical potential is analyzed.

The appropriate mass-like term in this case is $M=m\Lambda_{m}+\Delta_{so}\Lambda_{so}$ where $\Lambda_m$ and $\Lambda_{SO}$ are defined in (\ref{M}) and (\ref{SOham}) respectively. \\

We begin by the case where the chemical potential is zero.
The first difficulty that arises is that we can no longer write the Green's function in the simple form (\ref{easyGreen}). In the simple case of the SO coupling all the fermion flavors had the same dispersion relation, $\omega(\mathbf{k})=\pm\sqrt{\mathbf{k}^2+ \Delta^{2}_{so}}$, and we were dealing with a multiflavor two band model. We only had an unique pole in the Green's function and the degeneracy of the system came from the matrix structure in the numerator of (\ref{easyGreen}). When considering the two terms, this degeneracy is partially lifted and we have a non-degenerated multiband model. It can be easily seen by diagonalizing the hamiltonian in the presence of the two terms. The four bands are given by the expressions $\omega=\pm\sqrt{\mathbf{k}^2+(m+\Delta_{so})^2}$ and $\omega=\pm\sqrt{\mathbf{k}^2+(m-\Delta_{so})^2}$. Effectively, our system is now made of two (doubly degenerated) two-band subsystems. However, although the denominator in (\ref{Green}) is not any
 more proportional to the identity matrix, it is still a diagonal matrix that can be inverted. Using expression (\ref{C5}) and following the recipe of the last section one can see that we have two copies of the SO problem with two different masses given by $m\pm\Delta_{so}$. The result is accordingly:
\begin{eqnarray}\label{stag+so}
C_{s}^{\mu\nu\rho} = e^2\dfrac{\epsilon^{\mu\nu\rho}}{4\pi}[\mathrm{sign}(\Delta_{so}-m)+\mathrm{sign}(\Delta_{so}+m)].
\end{eqnarray}
It is clear that the interplay between a staggered potential and a spin orbit coupling is such that the Chern-Simons coefficient is still quantized. When $\Delta_{so}=0$ we recover a topologically trivial insulator and when $m=0$ we recover the result of the previous section as expected. The new feature compared to previous results is that when both are non zero (and for $\vert\Delta\vert >\vert m\vert$) there is still a topological response of the system.\\
Lets turn now to the case of having a finite chemical potential, not discussed previously in the literature. Similarly to what happened in the intrinsic SO case, depending on the relative value of $\mu$ against $\Delta_{so}\pm m$ we will have different results. Without doing any extra work we can read the result from the considerations made in the case of the intrinsic spin-orbit coupling by changing the masses appropriately. There are 4 different cases:

\begin{itemize}
\item{$|\mu|<|\Delta_{so}-m|$
and $|\mu|<|\Delta_{so}+m|$}:

Under these conditions the chemical potential lies inside both gaps gap  and the response is still quantized:
\begin{eqnarray}
C_{s}^{\mu\nu\rho} = e^2\dfrac{\epsilon^{\mu\nu\rho}}{4\pi}(\mathrm{sign}(\Delta_{so}-m)+\mathrm{sign}(\Delta_{so}+m)).
\end{eqnarray}

\item {$|\mu|>|\Delta_{so}-m|$ and
$|\mu|>|\Delta_{so}+m|$}:\\

In this case, the chemical potential crosses the bands for both masses and hence both integrals give a non-quantized result analogous to (\ref{so+mu}):

\begin{eqnarray}
C_{s}^{\mu\nu\rho} = e^2\dfrac{\epsilon^{\mu\nu\rho}}{4\pi}\left( \dfrac{\Delta_{so}-m}{|\mu|}+\dfrac{\Delta_{so}+m}{|\mu|}\right) = e^2\dfrac{\epsilon^{\mu\nu\rho}}{2\pi}\left( \dfrac{\Delta_{so}}{|\mu|}\right) .
\end{eqnarray}

It is interesting to note that this result does not depend on the value of $m$. Finally, the two cases left  can be written in a compact way:

\item{$|\mu|>|\Delta_{so} \mp m|$ and $|\mu|<|\Delta_{so} \pm m|$}: \\
These are the cases where one integral gives a topological contribution, as the chemical potential lies inside one of the gaps, but the other subspace gives a nonquantized contribution as the chemical potential crosses one of its bands. Hence, the result is:

\begin{eqnarray}
C_{s}^{\mu\nu\rho} = e^2\dfrac{\epsilon^{\mu\nu\rho}}{4\pi}\left( \dfrac{\Delta_{so}\mp m}{|\mu|}+\mathrm{sign}(\Delta_{so}\pm m)\right).
\end{eqnarray}

The $\pm$ signs indicate wether it is the conduction or the valence band which is crossed.
\end{itemize}

From these simple analytic results one infers that there is a competition between the staggered potential and the intrinsic spin orbit, conditioned by the presence of the chemical potential. The inclusion of the finite chemical potential completes the known results for this case.

\subsection{Competition between a Rashba coupling and intrinsic spin-orbit coupling at finite chemical potential}
\label{sec_SO-Ra}}

The competition of the intrinsic SO coupling and the Rashba contribution was already studied in the original reference \cite{KM05} for zero chemical potential. As the Rashba term by itself does not open a gap in the spectrum it was found that the topological insulating phase only exists for absolute values of $\Delta_{so}$ bigger than $\vert\lambda_R\vert$. The inclusion of a finite chemical potential makes the physical analysis more interesting.
Unlike the other cases, this is more complex and a general analytical treatment with arbitrary $\mu$ is messy. A detailed calculation can be found in \cite{DDB09};  in what follows  we aim to complete their analysis with several new analytical results together with an interpretation of the divergences that appear in this case.
\subsubsection{The case $\mu=-\Delta_{so}$}
\begin{figure}
\begin{minipage}{.3\linewidth}
\begin{center}
\includegraphics[scale=0.35]{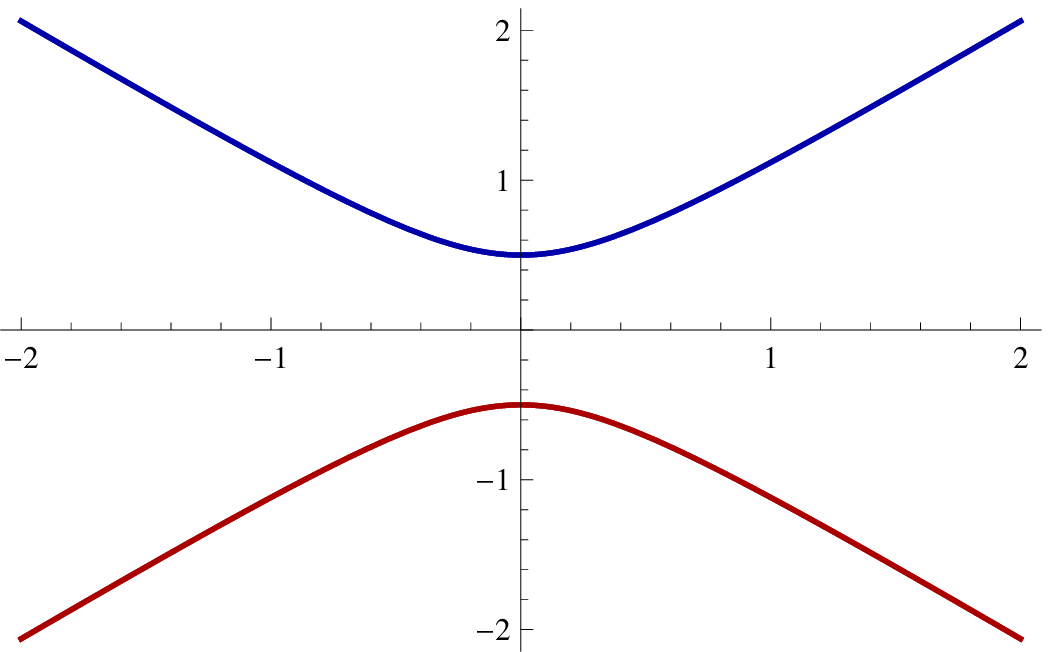}
\begin{center}
(a)
\end{center}
\end{center}
\end{minipage}
\begin{minipage}{.3\linewidth}
\begin{center}
\includegraphics[scale=0.35]{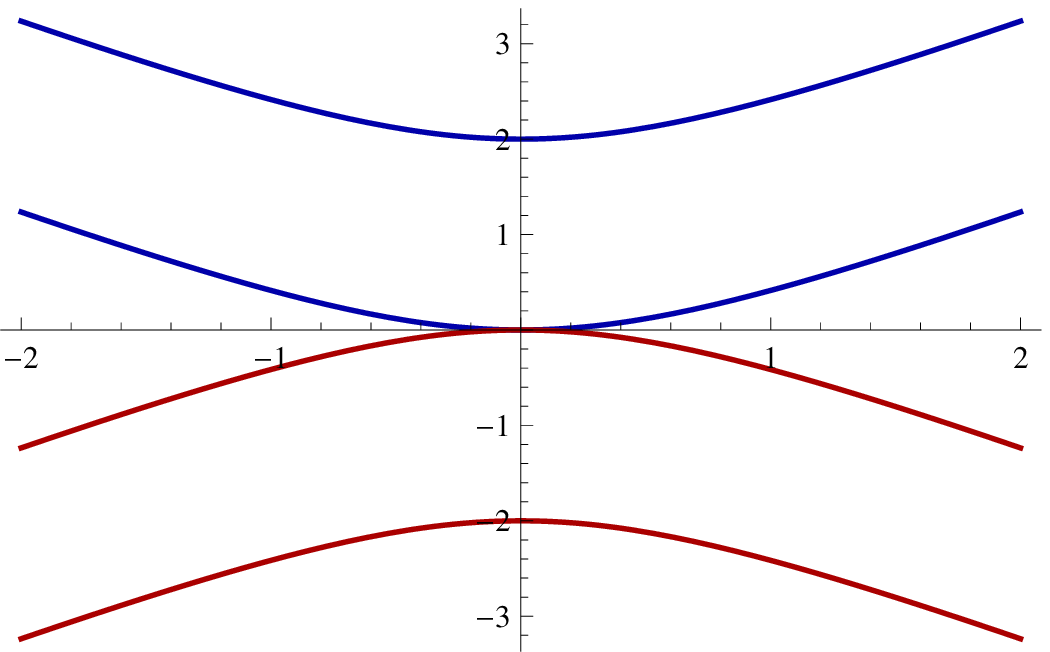}
\begin{center}
(b)
\end{center}
\end{center}
\end{minipage}
\begin{minipage}{.3\linewidth}
\begin{center}
\includegraphics[scale=0.35]{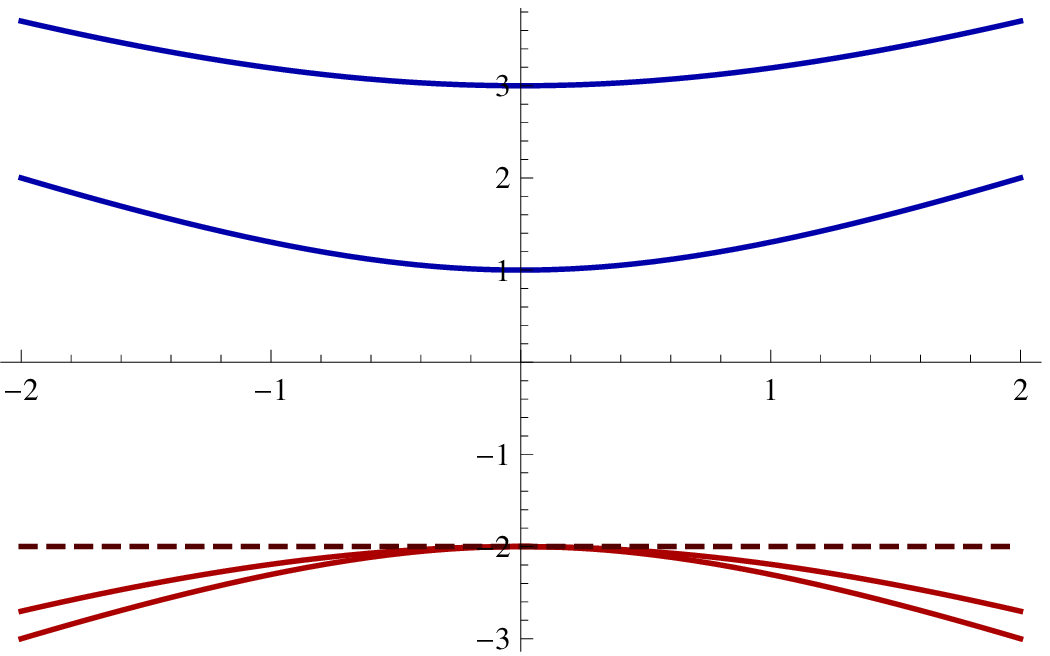}
\begin{center}
(c)
\end{center}
\end{center}
\end{minipage}\\
\begin{minipage}{.3\linewidth}
\begin{center}
\includegraphics[scale=0.35]{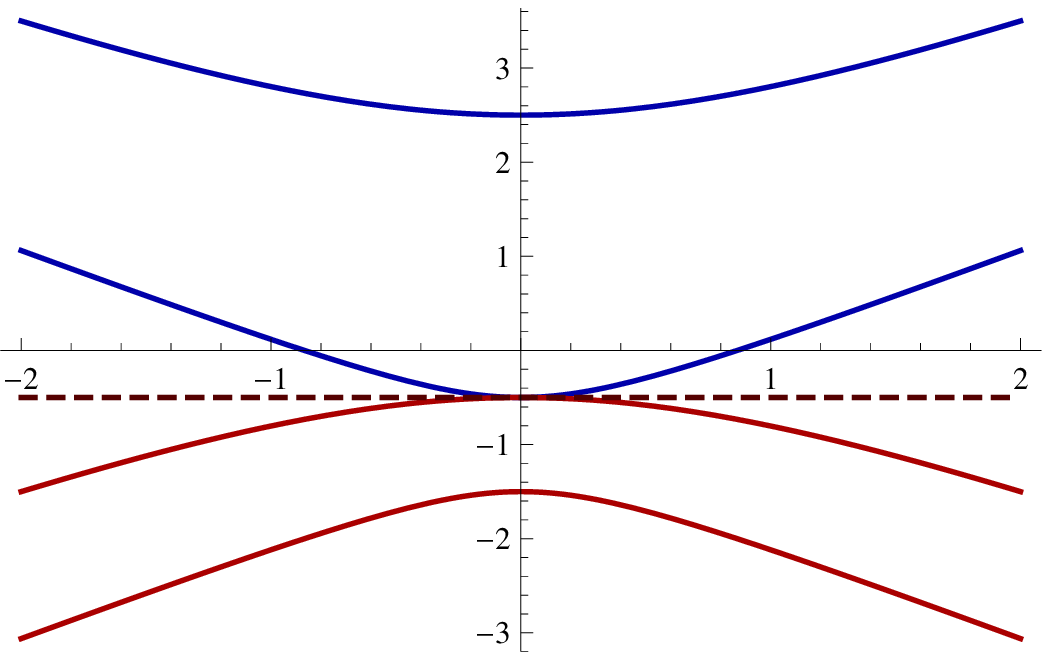}
\begin{center}
(d)
\end{center}
\end{center}
\end{minipage}
\begin{minipage}{.3\linewidth}
\begin{center}
\includegraphics[scale=0.35]{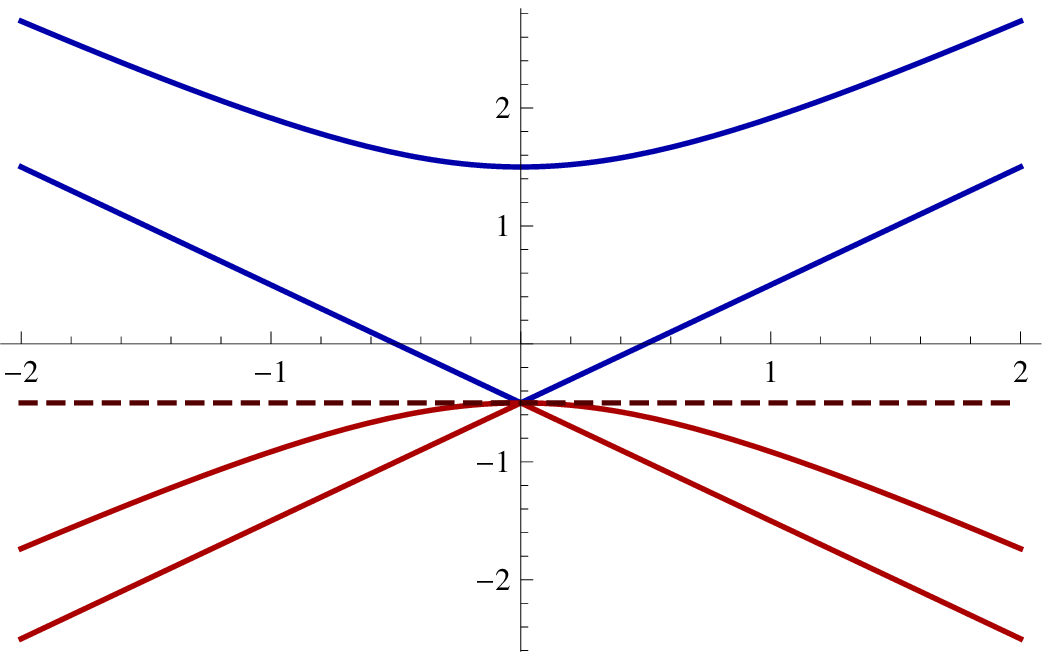}
\begin{center}
(e)
\end{center}
\end{center}
\end{minipage}\\
\caption{\label{Bandsrashso} Band structure for different values of $\lambda_{R}$ and $\Delta_{so}$. (a) With $\lambda_{R}=0$ and $\Delta_{so}\neq0$ a gap opens. The bands are degenerate in spin. (b) With $\lambda_{R}\neq0$ and $\Delta_{so}=0$ the gap is closed but the spin degeneracy is lifted. (c) With $\lambda_{R} < \Delta_{so}$ a gap opens in the spectrum.(d) With $\lambda_{R} > \Delta_{so}$ the gap closes. (e) When $\lambda_{R} = \Delta_{so}$ three of the bands touch which leads to a divergence in $C^{\mu\nu\rho}_{5}$ (see text). The horizontal solid line indicates where the chemical potential would lie if it was fixed at $\mu=-\Delta_{so}$. The band structure at this point generates a divergence when $\lambda_{R} > \Delta_{so}$. }
\end{figure}

The band structure when these couplings are included is shown in figure \ref{Bandsrashso}. As it is known in the presence of a Rashba term $\lambda_{R}$  the spin degree of freedom is no longer a good quantum number and we do not expect the spin hall conductivity to be quantized. After performing the trace in \eqref{C5}, the integral left is given by:

\begin{equation}\label{rash+so}
C_{s}^{\mu\nu\rho} = \dfrac{e^2}{2}\int \dfrac{dk^3}{(2\pi)^3}(-8i)\epsilon^{\mu\nu\rho}\frac{\left(4 \lambda_{R}^2 (k_{0}+\Delta_{so}) \left(k_{0}(k_{0}+\Delta_{so})-k_{x}^2+k_{y}^2\right)+\Delta_{so} \left(-k_{0}^2+\mathbf{k}^2+\Delta_{so}^2\right)^2\right)}{\left(\left(-k_{0}^2+\mathbf{k}^2+\Delta_{so}^2\right)^2-4 \lambda_{R}^2 (k_{0}+\Delta_{so})^2\right)^2},
\end{equation}
where $k^{\pm A}_{0}= \pm\sqrt{\mathbf{k}^2+(\lambda_{R}-\Delta_{so})^2}-\lambda_{R}-\mu$ and  $k^{\pm B}_{0}= \pm \sqrt{\mathbf{k}^2+(\lambda_{R}+\Delta_{so})^2}+\lambda_{R}-\mu$ are the poles. The simplest case (regarding its pole structure) is the case $\mu=-\Delta_{so}$ (see fig \ref{Bandsrashso}), since for all values of the parameters there are two bands below (and two above) the chemical potential, therefore defining the same pole structure for all values of $\mathbf{k}$. Since the chemical potential is placed at a special point of the band structure were the bands meet when the gap closes, we expect to have divergences which we interpret as the divergent contribution from the zero modes. \\
There are eight possible cases depending on the sign of $\lambda_{R}$, $\Delta_{so}$, $\Delta_{so}+\lambda_{R}$ and $\Delta_{so}-\lambda_{R}$. Since the integral in (\ref{rash+so}) does not distinguish the sign of $\lambda_{R}$ we can safely assume $\lambda_{R}>0$. This reduces the problem to four cases:

\begin{itemize}
\item{$\Delta_{so}>0$, $\Delta_{so}+\lambda_{R}>0$ and $\Delta_{so}-\lambda_{R}>0$:}\\
In these conditions  the gap remains open ($\Delta_{so}>\lambda_{R}$) we get a finite result:
\begin{equation}
C^{\mu\nu\rho}_{s} = e^2\dfrac{\epsilon^{\mu\nu\rho}}{4\pi}\dfrac{\Delta_{so}}{\lambda_{R}}\mathrm{ln}\bigg\vert\dfrac{\lambda_{R}+\Delta_{so}}{\lambda_{R}-\Delta_{so}}\bigg\vert.
\label{SOR}
\end{equation}

Note that when $\lambda_{R}/\Delta_{so}\rightarrow 0$ we can make use of the relation $$\mathrm{arctanh}(x)=\frac{1}{2}\ln\left(\frac{1+x}{1-x}\right),$$ to recover the standard result with the sign function obtained for $\Delta_{so}$ only.
\item{$\Delta_{so}<0$, $\Delta_{so}+\lambda_{R}<0$ and $\Delta_{so}-\lambda_{R}<0$:}\\
With an analogous analysis one arrives to the same expression but with a sign difference, which accounts for the change in sign of $\Delta_{so}$.
\item{$\Delta_{so}>0$, $\Delta_{so}+\lambda_{R}>0$ and $\Delta_{so}-\lambda_{R}<0$:}\\
In this case, one can see that the integral diverges. As explained above, we now have a semi metal with the chemical potential set at the touching point of the bands. The contribution of the $\mathbf{k}=0$ particles turns the integral to be divergent.
\item{$\Delta_{so}<0$, $\Delta_{so}+\lambda_{R}>0$ and $\Delta_{so}-\lambda_{R}<0$:}\\
Again, the integral is found to be divergent but with an opposite sign.
\end{itemize}

We can summarize the above result in two compact expressions. For the case when the gap is opened, i.e. $\Delta_{so} > \lambda_{R} > 0$ and $\Delta_{so} < -\lambda_{R} < 0$, one can write:

\begin{equation}
C^{\mu\nu\rho}_{s} = e^2\mathrm{sign}(\Delta_{so})\dfrac{\epsilon^{\mu\nu\rho}}{4\pi}\dfrac{\Delta_{so}}{\lambda_{R}}\mathrm{ln}\bigg\vert\dfrac{\lambda_{R}+\Delta_{so}}{\lambda_{R}-\Delta_{so}}\bigg\vert.
\end{equation}

For the other case ($\lambda_{R}>\Delta_{so}>-\lambda_{R}$) where the gap is closed the integral diverges as $C^{\mu\nu\rho}_{5}\rightarrow\mathrm{sign}(\Delta_{so})\infty$ as a result of the particular position of the chemical potential, where it encounters an infinite contribution from degenerate zero modes.

\subsubsection{The case $\mu=0$}

When $\mu=0$ it is immediate to notice that two of the bands can cross the fermi energy depending on the value of the parameters. Hence, the chemical potential lies inside the gap or crosses one of the bands, depending on the value of the parameters (See figure \ref{Bandsrashso}). Again the integral \eqref{rash+so} which we have to evaluate has poles at the dispersion relation. An important difference with respect to other cases is that  the poles have a non-trivial dependence on $\mathbf{k}$. We consider some illustrative cases:

\begin{itemize}
\item{$\Delta_{so}>0$, $2\lambda_R>\Delta_{so}>\lambda_R$:}\\
This corresponds to a situation where the gap is open but the chemical potential is such that it intersects one of the bands, hence being the system metallic. The result is given by:

\begin{equation}
C^{\mu\nu\rho}_{s}=e^2 \dfrac{\epsilon^{\mu\nu\rho}\Delta_{so}}{4\pi\lambda_{R}}\mathrm{ln}\bigg\vert\dfrac{\Delta_{so}}{2 (\lambda_{R}+\Delta_{so})}\bigg\vert\ .
\end{equation}

Interestingly, the case where $\Delta_{so}>0$, $\lambda_R>\Delta_{so}$ has the same result. In this case the chemical potential intersects one of the bands but the gap is closed. These results suggest that when the chemical potential intersects one of the bands and hence the system becomes metallic, it does not matter whether a gap is open or not below the chemical potential as both situations give the result shown above.

\item{$\Delta_{so}>0$, $\Delta_{so}>2\lambda_R$:}\\
In this case, the chemical potential falls inside the bulk gap and so this situation is reminiscent of the situation with $\mu=-\Delta_{so}$ when the gap is open. The result is the one given in (\ref{SOR}).

\end{itemize}

As expected, we recover the result for the previous section, but, since the chemical potential is not at a singular point we do not encounter divergences. The other cases left can be worked out in a similar fashion. Since the information they carry is redundant they will not be discussed further. \\
We have thus completed the analysis made in \cite{DDB09} by introducing several novel analytical results together with an interpretation on the origin of the divergences. The competition of $\Delta_{so}, \lambda_R$ and staggered potential at zero chemical potential was studied in \cite{KM05b} and we will not analyze it here.

\section{Bilayer Graphene}
\label{sec_bil}
\subsection{The model}

One can extend the previous analysis to bilayer graphene where in addition to the valley ($\boldsymbol{\tau}$), spin ($\bf{s}$) and pseudospin ($\boldsymbol{\sigma}$) degrees of freedom we have the layer degree of freedom ($\boldsymbol{\mu}$). The new degree of freedom enables a richer playground for topological phases. Remarkably, the effective action formalism allows to span these phases with little effort, as we shall discuss next.

We will consider the  bilayer graphene Lagrangean given by
\begin{equation}\label{barelagblg}
\mathcal{L}^{(b)}_{0} = \bar{\psi}(\gamma^{\mu}k_{\mu}-\sum_{i}\lambda_{i}\Lambda_{i})\psi ,
\end{equation}
where the bilinear couplings will be described below and the $\gamma$ matrices are now constructed in the space of valley, pseudospin, spin, and layer. Following the notation of the monolayer they are chosen to be $\gamma_{0}=1_{\tau}1_{s}1_{\mu}\sigma_{z}$, $\gamma_{1}=i\tau_{z}1_{s}1_{\mu}\sigma_{y}$, $\gamma_{2}=-i1_{\tau}1_{s}1_{\mu}\sigma_{x}$.\\

\noindent
The bilinear terms include the standard interlayer coupling given in our notation by
\beq
\lambda_t\Lambda_t=\frac{1}{2}t_\perp 1_s1_\tau i(\mu_y\sigma_x-\mu_x\sigma_y),
\eeq
and an external gate potential which is known to open a gap in the bilayer system and which is given by
\beq
\lambda_V\Lambda_V=V 1_s1_\tau\mu_z\sigma_z.
\eeq
In addition we shall consider two types of bilinear terms, the staggered potential and the various spin-orbit couplings given in \cite{G10}. The staggered potential is just the trivial extension of the monolayer. It is given  by $$\lambda^{(b)}_{m}\Lambda^{(b)}_{m} = m1_{\tau}1_{s}1_{\mu}1_{\sigma},$$ where the superscript $(b)$ denotes bilayer graphene.

Following \cite{G10} (see also \cite{MK10}),
 one can construct the following spin orbit terms (in the notation of \cite{KM05}):
\begin{eqnarray}\nonumber
\lambda^{(b)}_{so}\Lambda^{(b)}_{so} &=& \lambda_{1}\tau_{z}s_{z}1_{\mu}\sigma_{z}+\lambda_{2}\tau_{z}s_{z}\mu_{z}1_{\sigma}+
\lambda_{3}\left(1_{\tau}s_{x}\mu_{z}\sigma_{y}-\tau_{z}s_{y}\mu_{z}\sigma_{x}\right)+\\
\nonumber
&+& \lambda_{4}\left(1_{\tau}s_{x}\mu_{y}\sigma_{z}-\tau_{z}s_{y}\mu_{x}\sigma_{z}\right).
\end{eqnarray}
The first term is the one corresponding to the (monolayer) intrinsic spin orbit coupling. The last three terms, which involve $\mu$ are intrinsic to bilayer graphene where the ones proportional to $\lambda_{3}, \lambda_{4}$ are Rashba-like term. We will see that the effect of $\lambda_{1}$ and $\lambda_{2}$ is to open a non-trivial gap in the spectrum and so turning the system into a topological insulator with unprotected edge states. Rashba-like terms where studied numerically in \cite{LT10} and shown to have the same effect as in the monolayer, and hence we will not study their effect in this work.\\

The usual spin Chern-Simons term analogous to the one discussed in the monolayer case will be associated to the matrix
$$\gamma_{5}=1_{\tau}s_{z}1_{\mu}1_{\sigma}.$$
We will denote as before the value of the coefficient by $C_s$. Its computation follows the lines described in sect. \ref{sec_eff}. A non zero value indicates that the system is a spin topological insulator and the edges support spin currents.

As discussed in section \ref{sec_SCS} the present formalism allows us to describe other types of insulators by appropriately redefining $\gamma_5$ to resolve other degrees of freedom, which certifies the power of the method. In particular we will compute  a ``valley" Chern number to study the ``valley" topological insulator character  of the system. This insulator will be characterized by a matrix $$\gamma_{5,v}=\tau_{z}1_{s}1_{\mu}1_{\sigma},$$  tailored to resolve the valley degree of freedom. The corresponding Chern-Simons coefficient will be named $C_K$. A non zero value of this coefficient implies a topological insulator with valley polarized edges.

\subsection{Summary of the results for bilayer graphene:}

We have computed the $C_{s}$ terms for bilayer graphene along the same lines described previously for the case of the monolayer. We summarize the results in  Table 1 where the non zero couplings are listed ($t_{\perp}$ is always non zero and positive). As discussed above, we have also calculated the $C_{K}$ Chern number which is the ``valley" Chern number calculated with  $\gamma_{5}$ replaced by $\gamma_{5,v}$ in the trace \eqref{C5} and get similar results.  \\
We omit an explicit derivation of the results since the integrals appearing are similar to the monolayer case, which we have detailed earlier and no new technical issue appears for the case of the bilayer.
Nevertheless, the table also includes the technical detail of why the result vanishes in cases when it is zero, and the method of evaluation of the spatial integral if the result is finite. The chemical potential in all cases is set to zero, which in all cases falls inside the gap. With this set up, no band crosses the Fermi energy and the integrals are easily evaluated.

\hspace*{2cm}
\begin{center}
\begin{table}[hts]
\begin{tabular}{c|c|c}\label{table1}
 Non zero couplings & $C_{s}$ & $C_{K}$ \\
\hline\hline
 $ V\neq 0 $  & 0 (trace vanishes) & $\frac{2e^2}{2\pi}\mathrm{sgn(V)}$ (numerical)\\
\hline
 $\lambda_{1}\neq 0$   & $\frac{2e^2}{2\pi}\mathrm{sgn(\lambda_{1})}$ (analytic) & 0 (trace vanishes)\\
 \hline
 $\lambda_{2}\neq 0 $   & $\frac{2e^2}{2\pi}\mathrm{sgn(\lambda_{2})}$ (numerical) & 0 (trace vanishes)\\
 \hline
 $\lambda_{3}\neq 0 $   & 0 (vanishes by parity) & 0 (trace vanishes) \\
 \hline
 $\lambda_{4}\neq 0 $   & 0 (trace vanishes) & 0 (trace vanishes)\\
 \hline
  $\lambda_{1}\neq 0 $  $\lambda_{2}\neq 0 $    & $\frac{2e^2}{2\pi}\mathrm{sgn(\lambda_{1}+\lambda_{2})}$ (numerical) & 0 (trace vanishes)\\
 \hline
  $\lambda_{1}\neq 0 $  $V\neq 0 $    & $\frac{e^2}{2\pi}(\mathrm{sgn}(\lambda_{1}+V)+\mathrm{sgn}(\lambda_{1}-V))$ (numerical) & $\frac{e^2}{2\pi}(\mathrm{sgn}(V+\lambda_{1})+\mathrm{sgn}(V-\lambda_{1}))$ (numerical)\\
 \hline
 $\lambda_{2}\neq 0 $  $V\neq 0 $    & $\frac{e^2}{2\pi}(\mathrm{sgn}(\lambda_{2}+V)+\mathrm{sgn}(\lambda_{2}-V))$ (numerical) & $\frac{e^2}{2\pi}(\mathrm{sgn}(V+\lambda_{2})+\mathrm{sgn}(V-\lambda_{2}))$ (numerical)\\
 \hline
  $m\neq 0 $   & 0 (trace vanishes) & $\frac{2e^2}{2\pi}\mathrm{sgn(m)}$ (analytic)\\
 \hline
\end{tabular}
\caption{Classification of the two topological insulating phases of bilayer graphene considered in the text.}
\end{table}
\end{center}

These results indicate that bilayer graphene can be a ``valley" topological insulator, as it has been  recently suggested \cite{M10} by only  applying a gate voltage $V$  and also a topological insulator when $\lambda_{1}$ and $\lambda_{2}$ -- which open non trivial gaps -- are present. However, note that the Chern number for bilayer graphene is two times bigger than for the case of the monolayer in all cases. This even Chern number implies that we are dealing with a topological insulator with unprotected edge states. The result that bilayer graphene is a topological insulator for $\lambda_{1}$ was already mentioned in \cite{G10} but nothing was said about $\lambda_{2}$ which also seems to make the bilayer system a topological insulator.\\
The origin of a non-trivial gap given by $\lambda_{2}$ can be traced back to the matrix form of this coupling, together with the form of a $t_{\perp}$ coupling in bilayer graphene and the structure  of $\gamma_{5}$. All these together provide the complete set of matrices in the layer index ($\mu_{x},\mu_{y},\mu_{z},1_{\mu}$) so that the trace does not vanish and a non trivial spin Chern-Simons is generated, having however a ``layer-like" origin.
It is easy to check that when $t_{\perp}=0$,  $\lambda_{2}$ does not contribute to the spin Chern-Simons term. In contrast, $\lambda_{1}$ in this situation still does contribute, indicating that the coupling between layers is of critical importance for this spin-orbit like coupling to have an effect.

From the table of results it is clear that the combined perturbation $\lambda_{1}$ and $\lambda_{2}$ can enhance the topological response. The reason why the couplings $\lambda_{1}$ and $\lambda_{2}$ have the same effect can be understood in terms of the effective low energy hamiltonian. The equivalence can be traced back to a symmetry analysis of the effective model performed in ref. \cite{MK10}. This analysis shows that the only term allowed by time reversal and discrete spatial symmetries is the Kane-Mele term associated to $\lambda_{1}$. It means that although $\lambda_{1}$ and $\lambda_{2}$ have a different microscopic origin, both couplings will lead to the same term in the low energy approximation explaining why in table I the only effect of $\lambda_{2}$ is to renormalize $\lambda_{1}$.

The case with $\lambda_{i}\neq 0$ ($i=1,2$) and $V\neq 0$ is analogous to the monolayer case where the staggered potential competes with the spin orbit coupling. As expected, the trivial coupling is $V$ when calculating $C_{s}$ and  $\lambda_{i}$ when calculating $C_{K}$. This indicates that the (valley) topological nature of the bilayer is affected by the presence of the gate potential $V$.

As mentioned above, the Rashba like terms $\lambda_{3}$ and $\lambda_{4}$ go against the topological nature of bilayer graphene as it was discussed in \cite{LT10}.
Table 1 shows that bilayer graphene is a rich playground to understand the competition between different topological phases. The addition of a finite chemical potential will change this picture as it happens in the monolayer case.
It is worth to mention that the computation of the different terms in table I is similar in many aspects to the computation of $C_{s}$ in the case of monolayer graphene in the presence of  $\lambda_{R}+\Delta_{so}$ couplings. In the present case the existence of the perpendicular hopping term $t_{\perp}$ makes the low energy hamiltonian quadratic in the momentum operator and $C_{s}$ will have a form similar to the expression (\ref{rash+so}) with the obvious modifications concerning the position of the Fermi level inside the gap.

\section{Conclusions and open questions}
\label{sec_final}
We have provided a unifying view to generate and compute different Chern-Simons terms that can appear in mono and bilayer graphene, and used the formalism to study various topological insulating phases in the system.  Part of the results of this work were already  obtained in the literature and others are distinct. In particular in the monolayer case we have studied the competition of the intrinsic spin orbit term with a staggered potential in the presence of finite chemical potential, and we have completed the analysis of \cite{DDB09} by obtaining an analytical expression for the Chern-Simons term in the case of having an intrinsic  SO and a Rashba term at finite chemical potential. We have also clarified the origin of the divergences encountered for the particular case of a chemical potential equal to the intrinsic SO value.  For the bilayer graphene we have characterized the usual spin insulator and a different non-trivial topologically insulating phase similar to the one described in \cite{M10}, that can be thought of as a valley topological insulator, under the same formalism. The topological nature of the valley insulator was shown to be affected by the introduction of a gate voltage. We have also seen that the layer degree of freedom does not give rise to topological insulating phases with the physical couplings discussed in this work. Obviously the formal identity of the Hamiltonians with various couplings allows to play with possible couplings to trade spin by layer degrees of freedom. The major  interest will lie in finding couplings that will be experimentally realizable and can lead to non-trivial applications as discussed in \cite{M10}, an issue out of the scope of the present work.

The effective action formalism used in this work allows to compute   non quantized values for the spin Chern numbers, in contrast to other methods, like the computation based on the Berry phase, which only gives the quantized parts. This non-quantized Chern number is still linked to a spin Hall effect which means that we still have counter propagating edge states whose spin polarization is given by this Chern number. A non quantized value means that the spin (or generally speaking the quantum number carried by the propagating edge state) is not strictly conserved. In spite of this, there have been some theoretical studies (in the case of valley Hall effect) claiming that although  not  protected by symmetries, these counter propagating edge currents  are quite robust against the effect of disorder and the coherence length of the quantum number they carry can be significantly large to be used in some electronic or spintronic applications \cite{M10}.

\section{Acknowledgments}
We thank A. Morpurgo for sharing with us unpublished results. A.G.G. acknowledges very useful discussions with F. de Juan. A. C. has enjoyed conversations with E. McCann and is  supported by EPRSC Science and Innovation award EP/G035954. Support by MICINN (Spain) through grant FIS2008-00124 is acknowledged.

\bibliography{SCS8}

\end{document}